\documentclass[12pt,hidelinks]{article}

\usepackage{siunitx}   
\usepackage[version=4]{mhchem}    
\usepackage{hyperref}
\usepackage{scicite}
\usepackage{graphicx}
\usepackage{upgreek}
\usepackage{caption}
\usepackage{multirow}
\usepackage{subcaption}
\usepackage{gensymb}
\usepackage{booktabs}
\usepackage{rotating}
\usepackage{array}
\usepackage{tabularx}
\usepackage{float}

\usepackage{makecell}

\usepackage[export]{adjustbox}

\newcolumntype{Y}{>{\centering\arraybackslash}X}
\usepackage[dvipsnames]{xcolor}

\newcolumntype{C}{>{\centering\arraybackslash}m{2.5cm}}
\newcolumntype{M}[1]{>{\centering\arraybackslash}m{#1}}

\usepackage{times}

\newenvironment{sciabstract}{%
\begin{quote} \bf}
{\end{quote}}

\usepackage{amsmath,amssymb}
\usepackage[version=4]{mhchem} 
\usepackage{siunitx}           
\title{Channel-last gate-all-around nanosheet oxide semiconductor transistors}

\author
{Fabia F. Athena$^{1,*,\dagger}$, Xiangjin Wu$^{1,\dagger}$, Nathaniel S. Safron$^{2,\dagger}$, \\ Amy Siobhan McKeown-Green$^{3}$, Mauro Dossena$^{4}$, Jack C. Evans$^{1}$, \\
Jonathan Hartanto$^{1,5}$, Yukio Cho$^{6,7}$,
Donglai Zhong$^{2}$, Tara Peña$^{1}$, \\ Paweł Czaja$^{6,8}$,  Parivash Moradifar$^{6}$, Paul C. McIntyre$^{5}$,\\
Mathieu Luisier$^{4}$, Yi Cui$^{5}$, Jennifer A. Dionne$^{5}$, Greg Pitner$^{2}$, 
\\ Iuliana P. Radu$^{9}$, Eric Pop$^{1}$, Alberto Salleo$^{5}$, \\H.-S. Philip Wong$^{1,*}$\\
\\
\normalsize{$^{1}$Department of Electrical Engineering, Stanford University, Stanford, CA, USA}\\
\normalsize{$^{2}$Corporate Research, Taiwan Semiconductor Manufacturing Company, Ltd., San Jose, CA, USA}\\
\normalsize{$^{3}$Department of Chemistry, Stanford University, Stanford, CA, USA}\\
\normalsize{$^{4}$Department of Information Technology and Electrical Engineering, ETH Zürich, Switzerland}\\
\normalsize{$^{5}$Department of Material Science and Engineering, Stanford University, Stanford, CA, USA}\\
\normalsize{$^{6}$Department of Chemical Engineering, Stanford University, Stanford, CA, USA}\\
\normalsize{$^{7}$Applied Energy Division, SLAC National Accelerator Laboratory, Menlo Park, CA, USA}\\
\normalsize{$^{8}$Institute of Metallurgy and Materials Science, Polish Academy of Sciences, Poland}\\
\normalsize{$^{9}$Corporate Research, Taiwan Semiconductor Manufacturing Company, Ltd., Hsinchu, Taiwan}\\
 {\normalsize $^{*}$\textit{fathena@stanford.edu}, {\normalsize \textit{hspwong@stanford.edu}}}
{\normalsize $^{\dagger}$\textit{equal contribution}}
}

\date{} 
\begin{document}
\baselineskip24pt
\maketitle
\begin{sciabstract}
As we move beyond the era of transistor miniaturization, back-end-of-line-compatible transistors that can be stacked monolithically in the third dimension promise improved performance for low-power electronics. In advanced transistor architectures, such as gate‑all‑around nanosheets, the conventional channel-first process involves depositing dielectrics directly onto the channel. Atomic layer deposition of gate dielectrics on back-end-of-line compatible channel materials, such as amorphous oxide semiconductors, can induce defects or cause structural modifications that degrade electrical performance. While post-deposition annealing can partially repair this damage, it often degrades other device metrics. We report a novel channel‑last concept that prevents such damage. Channel‑last gate‑all-around self‑aligned transistors with amorphous oxide‑semiconductor channels exhibit high on-state current ($>$ 1 mA/µm) and low subthreshold swing (minimum of 63 mV/dec) without the need for post‑deposition processing. This approach offers a general, scalable pathway for transistors with atomic layer deposited channel materials, enabling the future of low-power three-dimensional electronics.
\end{sciabstract}

\section*{Main}
For half a century, progress in microelectronics has been driven by planar two‑dimensional (2D) scaling, shrinking lithographic dimensions, and reducing transistor gate lengths. That trajectory is now constrained by fundamental physical limits\cite{franklin2022carbon,woods2024death}. According to the International Roadmap for Devices and Systems (IRDS), density growth at an acceptable energy and cost requires a transition from 2D scaling to three‑dimensional integration and system‑technology co‑optimization, with monolithic three‑dimensional (M3D) integration as a central pillar to sustain device density growth at acceptable energy and cost \cite{IRDS2022,IRDS2023MM}. In M3D, device layers are fabricated sequentially within the back-end-of-line (BEOL), which is the stage of semiconductor manufacturing during which metal interconnect layers are formed on top of the devices. Once the base silicon transistor layer is complete, subsequent BEOL tiers must be processed under a limited thermal budgets and with strict chemical constraints to prevent degradation of the underlying transistors and interconnects above \cite{Mallik2017IEDM}.

\begin{figure}[!h]
  \centering
  \includegraphics[width=\linewidth]{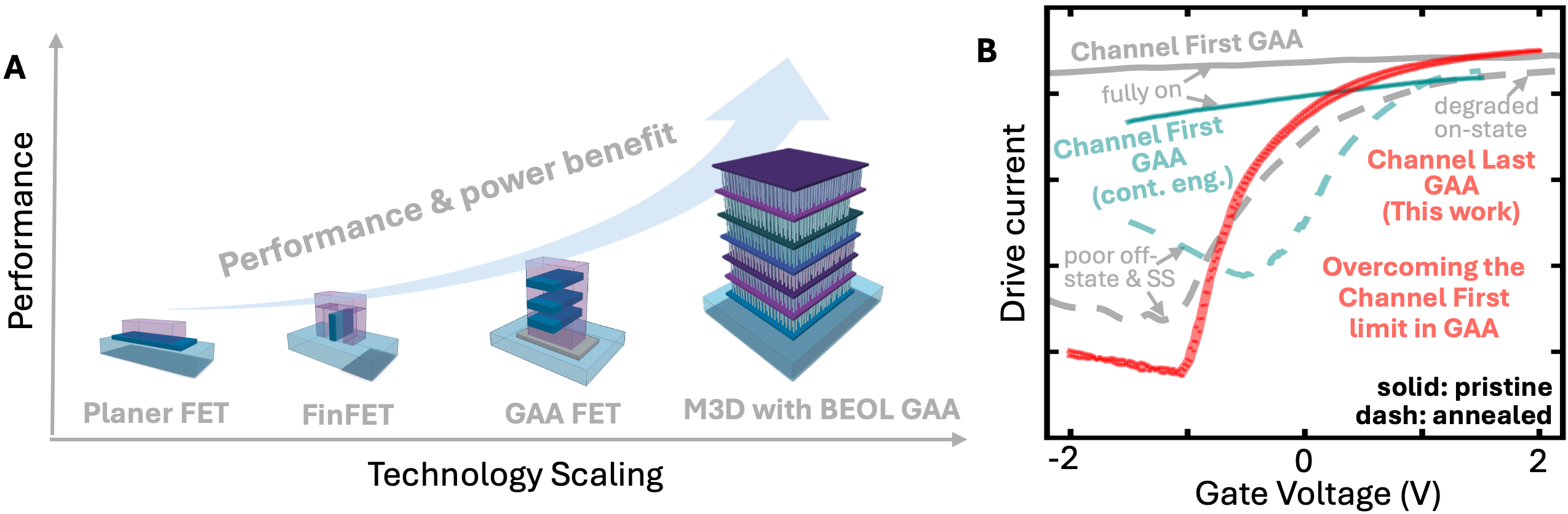}
  \caption{\textbf{Technology evolution with BEOL GAA transistors}. (\textbf{A}) Technology‑node scaling has progressed from planar to FinFET to GAA to dense M3D integration with GAA. BEOL compatible AOS based self‑aligned GAA transistors are attractive for M3D because they can be vertically integrated into dense multilayer stacks with improved electrostatics. (\textbf{B}) In AOS‑based GAA\cite{sarkar2025first}, traditionally the AOS channel is deposited first, followed by dielectric deposition directly on the channel (“channel‑first”). This process can damage the channel, adding additional oxygen vacancy donors thus, degrade device performance. Post‑deposition annealing is typically performed to recover the on/off ratio. However, it adds process complexity and can not fully recover the device metrices. Achieving high performance GAA with the traditional “channel-first” approach remains a central challenge. The proposed "channel last" (CL) GAA FET leverages a replacement‑channel concept that combines the benefits of BEOL‑compatible, back‑gated FETs and GAA architectures. This opens a path to M3D‑compatible GAA transistors built with ALD deposited any channel materials, such as AOS, 2D semiconductors.
}
  \label{fig:process}
\end{figure}

Low‑temperature deposition, and uniform, conformal coverage over high aspect ratio topography make atomic layer deposition (ALD) channel materials attractive for BEOL logic and memory fabricated in M3D. Among candidate channel materials compatible with BEOL fabrication, amorphous oxide semiconductors (AOS) grown by ALD have achieved good performance in back‑gated transistors without thermal annealing \cite{Si2021TED_Emode,athena2024first}. In back-gate structures, the channel is deposited after the gate dielectric (channel-last methodology) thus, the channel is not degraded by gate dielectric deposited on top of it. However, the back‑gated transistor structure is not self‑aligned and has non-ideal electrostatic control\cite{kim2005extremely,kim2005off}. For practical M3D BEOL logic and memory, target device geometries include self-aligned top-gate, dual-gate, and ultimately gate-all-around (GAA) nanosheet structures, with increasing degree of electrostatic control \cite{li2021first,jonsson2018self} (Fig.~\ref{fig:process}A).

Achieving high‑performance AOS channel transistors in a top‑gate or GAA geometry remains challenging. In conventional channel-first process flows for these structures, a high‑$\kappa$ dielectric is deposited directly onto the AOS channel. During the dielectric deposition process, additional oxygen‑vacancy donors and interface states are created or structural modifications occur that leave transistors fully turned-on after fabrication in their pristine state \cite{si2021high,Wahid2023EDL} (Fig.~\ref{fig:process}B). For AOS and 2D channel materials, device performance (e.g., on-current, threshold voltage ($V_{\mathrm{T}}$), hysteresis, subthreshold swing (S.S.)) is strongly influenced by the vacancy concentration, microstructure, interface quality, and defects at the interface and in the bulk \cite{yan2025clean,fortunato2012oxide}. Thus, the traditional channel-first process, which is typically used for top gate and GAA structures with AOS transistors causes a significant performance degradation. Using low‑temperature dielectrics or post‑deposition treatments (e.g., annealing or ozone) can reduce or heal these defects. However, these post-deposition steps introduce trade‑offs in dielectric quality (lower $\kappa$), process complexity, and thermal and chemical constraints \cite{Charnas2024ADMA}. Oxygen or nitrogen anneals can also degrade mobility, S.S. and BEOL wiring \cite{sarkar2024first,jana2025key}. Crucially, prolonged oxygen or nitrogen‑rich anneals can also oxidize or destabilize BEOL metals, compromise diffusion barriers and low‑$k$ dielectrics, thus jeopardizing completed interconnects and other sensitive circuit elements, rendering such treatments fundamentally incompatible with M3D integration \cite{kelly2018annealing,tsuya1994oxygen}.

How can we attain back-gate-comparable integration in a M3D compatible structure? This question motivates a different integration philosophy. We report a self‑aligned, channel‑last process for top‑gate, dual‑gate, or GAA transistors. This approach, \textit{inspired by} the nature's wonder geode, leverages replacement channel to deposit the channel material onto the gate dielectric (e.g., \ce{HfO2}) that is already formed in GAA structure rather than following the traditional channel-first approach for GAA. The delicate AOS channel does not experience the ALD step, thereby avoiding degradation from the top dielectric deposition, instead, the AOS channel is deposited in situ on the dielectric, enabling a low defect concentration \cite{zywotko2017thermal}. As we show, our fabricated channel‑last gate‑all‑around field effect transistors (CL GAA FET) show excellent performance without additional post-deposition treatment, opening a practical path to BEOL GAA logic and memory in M3D.

\section*{Channel-Last GAA Transistor Fabrication}

The fabrication process flow for the CL GAA FET was first formulated using process simulations using Coventor SEMulator3D~\cite{franke2017epe} followed by experimental fabrication. The important steps involved in the process flow for the CL GAA FET are shown in (Fig.~\ref{fig:tem}A). The process flow involves first patterning with electron beam lithography (EBL) and then depositing a local back gate and a sacrificial silicon dioxide (\ce{SiO2}) nanosheet. A metal anchor is then patterned and deposited to support the nanosheet structure in the channel width direction. After the gate anchor is in place, the sacrificial \ce{SiO2} is selectively removed by vapor phase etching at room temperature, which offers effectively nearly near-infinite selectivity ($>$ 1000: 1) over Ni\cite{witvrouw2000comparison,lee1997dry}. This etch step hollows out the space previously occupied by the nanosheet, creating a nanoscopic cavity. The cavity is then sequentially filled using plasma enhanced ALD (PEALD), first with a conformal layer of 10 nm thick \ce{HfO2} is deposited along the inner walls of the cavity forming the gate dielectric. Without breaking the vacuum, a 9 nm thick AOS layer of indium tungsten oxide (IWO) with 2\% W is then deposited to fill the remaining cavity without a seam using an optimized PEALD recipe \cite{yoo2024atomic,lin2024enhancement}, forming the transistor channel. 

Importantly, both the \ce{HfO2} and IWO depositions are carried out in-situ at 225 \degree C within the same ALD chamber without breaking the vacuum, ensuring a pristine, low-defect interface between the dielectric and channel. Finally, a self-aligned dry etching process is then performed to remove the channel material outside of the active region \cite{liang1994self,lukens2018self}. Details of the fabrication process are provided in the Materials and Methods in the Supplementary Materials (SM) (fig.~S1). Notably, our process is inspired by the principles of a natural geode formation. For CL GAA FETs (geode), a nanocavity (void) is first created in the gate metal (the outer shell). The cavity is then conformally lined with a high-k dielectric (the mineral lining), and finally filled with an AOS deposited last, which serves as the channel (the inner infill).

\section*{Structural Characterization}

\begin{figure}[!h]
  \centering
  \includegraphics[width=\linewidth]{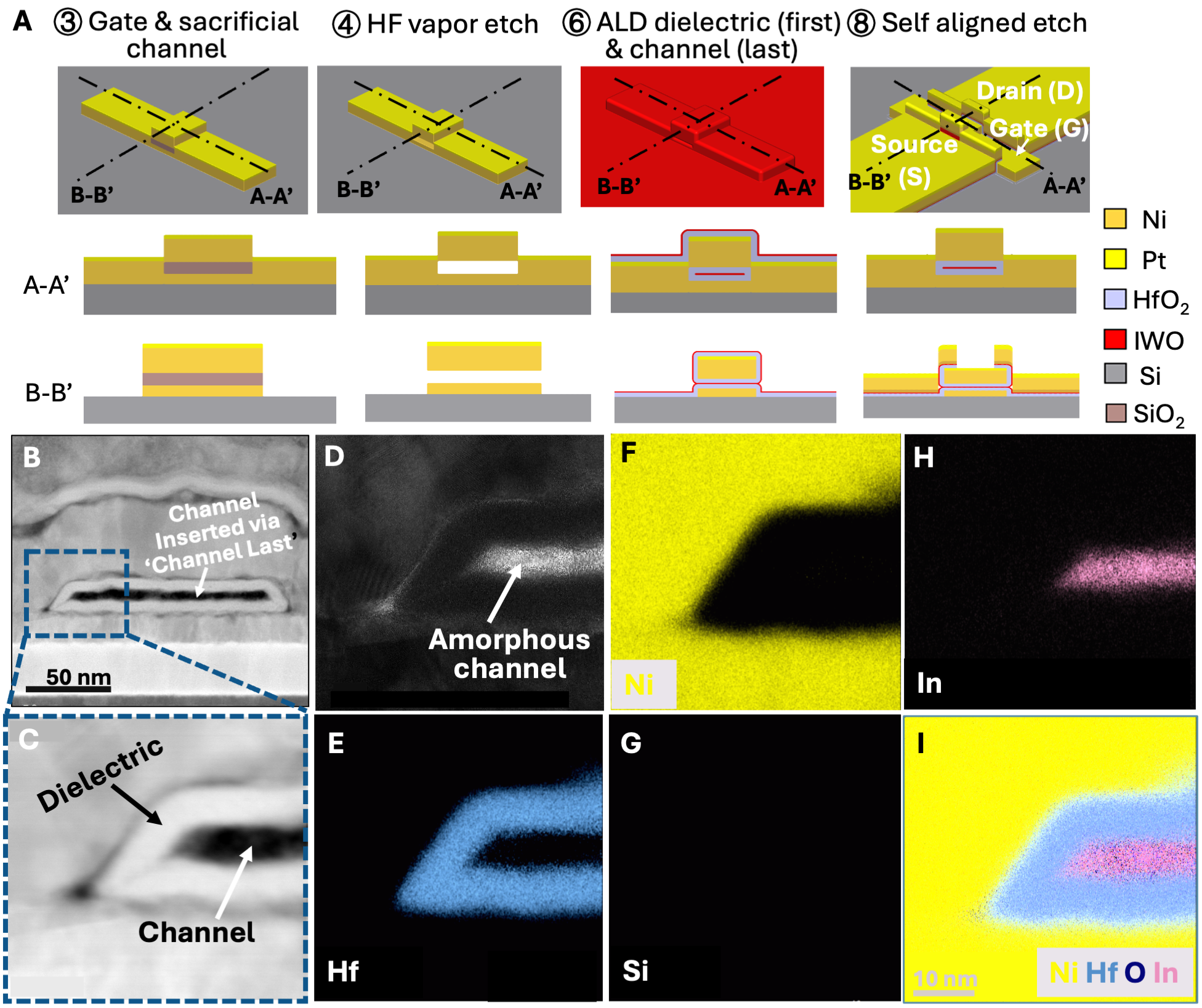}

    \caption{\textbf{Channel-last gate-all-around field-effect transistor (CL GAA FET).} The channel-last (CL) approach extends the channel-last advantages of back-gated FETs to GAA architectures. (\textbf{A}) Process flow of the CL GAA FET. A silicon dioxide (\ce{SiO2}) nanosheet and a local back gate are deposited and patterned, after which a \ce{Ni} anchor is formed to support the nanosheet. HF vapor selectively removes the \ce{SiO2} with near-infinite selectivity ($>$ 1000: 1) over Ni, creating a cavity that is sequentially filled by ALD, first with a hafnium oxide (\ce{HfO2}) gate dielectric, then with an indium tungsten oxide AOS channel. (\textbf{B}) Cross sectional bright-field TEM image of the fabricated CL GAA FET. (\textbf{C}) Zoomed view of the cross-sectional TEM image. (\textbf{D}) HAADF STEM image showing the amorphous channel within the gate dielectric. (\textbf{E})--(\textbf{I}) EDS elemental maps of the highlighted region, confirming that the AOS channel is surrounded by \ce{HfO2} dielectric, consistent with the channel-last GAA architecture.}

  \label{fig:tem}
\end{figure}

Cross-sectional transmission electron microscopy (TEM) was performed to verify the structure of the CL GAA FET. The cross-sectional bright-field TEM image (Fig.~\ref{fig:tem}B) reveals the gate stack and channel morphology. The ALD-deposited channel is clearly visible as a uniform layer within the \ce{HfO2}-lined cavity, indicating successful in-cavity deposition via the channel-last process. A magnified view of the left side of the CL GAA FET (Fig.~\ref{fig:tem}C) highlights the \ce{HfO2} gate dielectric and the IWO AOS channel residing at the core. Both layers appear continuous and uniform. Their measured thicknesses of $\sim$10 nm for \ce{HfO2} and $\sim$9 nm for IWO match the design targets. This confirms the cavity has been evenly filled with the desired thickness \cite{elers2006film}. The cross sectional TEM along the B-B' cut, shown in the SM (fig.~S2 ), also indicates that self aligned dry etching was successful. The high-angle annular dark-field (HAADF) TEM micrograph shows the AOS channel inside the channel-last architecture (Fig.~\ref{fig:tem}D). 

Energy-dispersive X-ray spectroscopy (EDS) elemental maps were collected over the magnified region shown in the bright-field image to confirm the elemental composition and spatial distribution of each layer. The Ni gate metal surrounds the structure as intended, the \ce{HfO2} dielectric uniformly coats the inner surface of the Ni gate, and the IWO channel material occupies the center of the structure. No Si signal is observed, confirming that the sacrificial \ce{SiO2} removal process was successful. In the overlaid EDS maps illustrating that the channel is fully enclosed by the dielectric and gate on all sides. The EDS analysis, together with the TEM imaging, conclusively shows that the channel-last integration approach successfully incorporated the AOS channel within the cavity, achieving the intended GAA structure.

\section*{Electrical Characterization}
An scanning electron microscope (SEM) image of a representative device structure are shown in Fig.~\ref{fig:elec}A. The atomic force microscope (AFM) image in Fig.~\ref{fig:elec}B confirms that the PEALD grown IWO channel (deposited on a blanket sample) has a relatively smooth surface (root-mean-square surface roughness $R_\mathrm{q} <450 \pm20$ pm). Direct current (d.c.) transfer characteristics ($I_{\mathrm{D}}$ versus $V_{\mathrm{GS}}$) of a CL GAA FET measured at $V_{\mathrm{DS}} = 0.05~\text{V}$ and $1.0~\text{V}$ (Fig.~\ref{fig:elec}C) exhibit a high drive current and an on/off current ratio exceeding $10^{8}$. Here, the channel thickness ($t_\mathrm{ch}$) is 6 nm, the channel length ($L$) is 185 nm and the width ($W$) is 75 nm. The on-state current is $>$1 mA/µm at $V_\mathrm{GS} = 2.0~\text{V}$ and $V_\mathrm{DS} = 1.0~\text{V}$. The current is normalized with the footprint width. The $V_{\mathrm{T}}$ extracted by the constant current method (at 100 nA*W/L) is around $-0.5~\text{V}$, which is consistent with the $V_{\mathrm{T}}$ of a back-gated device with a ~3 nm AOS channel (fig.~S3), indicating a $V_{\mathrm{T}}$ that is comparable to a back-gate OSFET can be achieved because the channel is deposited after the gate dielectric in both cases.


The $I_{\mathrm{D}}$ versus $V_{\mathrm{GS}}$ characteristics of a CL GAA FET device with $t_\mathrm{ch}$ = 9 nm, $L$ = 185 nm and $W$ = 75 nm are shown in Fig.~\ref{fig:elec}D. The device shows high $I_\mathrm{on}$ of 1.12 mA/µm at $V_\mathrm{GS} = 2.0~\text{V}$ and $V_\mathrm{DS} = 1.0~\text{V}$. The field effect mobility, extracted from the transfer curves (Fig.~\ref{fig:elec}E), reaches over 27 cm$^2$/(V·s) at $V_\mathrm{DS} = 0.05~\text{V}$. Furthermore, the S.S. is low, with a minimum of $\sim$63 mV/dec and an average over one decade is 65.9 mV/dec(Fig.~\ref{fig:elec}F). This sharp turn-on is consistent with a high-quality dielectric/AOS interface\cite{lee2016subthreshold,jiang2019printed,lyu1993determination}, as expected from the in situ channel-last process. Output characteristics ($I_{\mathrm{D}}$ versus $V_{\mathrm{DS}}$) for gate biases incremented in 0.5 V steps are shown in Fig.~\ref{fig:elec}G. The device delivers a maximum $I_\mathrm{on}$ of approximately 1 mA/µm at $V_{\mathrm{DS}} = 1.0~\text{V}$ and $V_{\mathrm{GS}} = 2.0~\text{V}$. Additionally, a device with a thinner ($\sim$3 nm) AOS channel (transfer curves shown in Fig.~\ref{fig:elec}H) exhibits a positive $V_{\mathrm{T}}$ of about $0.43~\text{V}$. $V_{\mathrm{T}}$ in AOS CL GAA devices can be further fine-tuned orthogonally by using $V_{\mathrm{T}}$ tuning approach that introduces a dipole layer\cite{athena2025orthogonal}.

To validate the measurements, we performed three‑dimensional Technology Computer-Aided Design (TCAD) simulations of a CL GAA FET. The model was calibrated to the \(t_{\mathrm{ch}}\) = 9\,nm device by matching the \(I_{\mathrm{D}}{-}V_{\mathrm{GS}}\) characteristics at \(V_{\mathrm{DS}}=1 V\). Within physically reasonable bounds, we adjusted the gate‑metal work function to reproduce $V_{\mathrm{T}}$  and the S.S., and tuned the low‑field mobility to match the on‑current. The geometry of the cut plan is shown in fig~S4. The fitting shows good agreement with the experimental data (fig.~S5). Fig.~\ref{fig:elec}I shows the simulated electron density (cm\textsuperscript{-3}) distribution in the nanosheet cross-section. As expected for a GAA geometry, the electron density is high along the entire nanosheet periphery in the on state and strongly suppressed throughout the nanosheet in the off state, indicating uniform depletion. This demonstrates that the GAA geometry provides improved off‑state electrostatic control of the channel. Figure~\ref{fig:elec}J compares simulated and experimental $V_{\mathrm{T}}$ , showing that $V_{\mathrm{T}}$ increases as \(t_{\mathrm{ch}}\) is reduced, as expected for AOS FETs. Additional TCAD fitting data, parameter values, electric field \(|\mathbf{E}|\)  distribution and sensitivity sweeps are provided in SM (fig.~S6).

Finally, we benchmarked the performance of CL GAA FET ($t_\mathrm{ch}$= 9 nm) against other reported AOS transistor architectures. Our CL GAA FET demonstrates a higher on-current, a steeper S.S., and a greater field-effect mobility compared to these prior works \cite{Wahid2023EDL, sarkar2024first, han2021first, zhang2022a, li2023first, lu2022performance, chen2022scaling, ye2020double, chakraborty2020beol, ye2025high, tong2025boosted, zhao2024first, yang2025performance, gu2023high, luo2025the}, as shown in Figs.~\ref{fig:elec}K--N. Notably, compared to other efforts, our process incorporates self‑aligned steps that eliminate lithographic overlay error between critical device layers. This is particularly important for small gate‑length transistors, because it reduces the parasitic capacitance arising from unintended overlap between layers. Moreover, the fabrication process is entirely low-temperature, the maximum process temperature is only 225\degree C, and no additional post-deposition annealing or ozone treatment is required.

\section*{Role of Deposition Order} 

To assess the effect of deposition order on defects and the structure of the oxide semiconductor channel, we analyzed indium oxide-hafnium oxide samples in two stacking sequences: (i) \ce{HfO2} deposited first, followed by IWO (channel-last), and (ii) IWO deposited first, followed by \ce{HfO2} (channel-first) deposited on a blanket \ce{SiO2} substrate. For both stacks, the IWO layers were grown within the same ALD deposition run to avoid any deposition-related variations and to ensure a fair comparison. Cross-sectional HAADF-STEM combined with STEM-EDS was used for structural and compositional characterization. Fig.~\ref{fig:dft}A shows a HAADF-STEM image of the channel-last stack. The interface is uniform across the stack, with no variation observed. Fig.~\ref{fig:dft}B presents EDS maps of O, In, Hf, and W from a representative region. Fig.~\ref{fig:dft}C shows a high-resolution BF-STEM image along the \ce{SiO2} [110] zone axis with an inset FFT, showing the amorphous nature of the IWO layer, selected regions of interest (ROIs) are indicated for FFT analysis. Fig.~\ref{fig:dft}(D-E) FFTs from the ROIs in Fig.~\ref{fig:dft}C exhibit diffuse amorphous rings without pronounced diffraction spots, confirming that the indium oxide layer is fully amorphous.
Fig.~\ref{fig:dft}F shows the HAADF-STEM image of the channel-first stack. Some dark spots at the interface between IWO and \ce{HfO2} are observed (indicated with a guide arrow), consistent with non-uniformity or interfacial defects. The corresponding X-EDS elemental mapping (Fig.~\ref{fig:dft}G) confirms the chemical composition and distribution of all layers. Fig.~\ref{fig:dft}H shows a lattice-resolved BF-STEM image of the channel-first film along the \ce{SiO2} [110] zone axis (inset FFT), with ROIs in the IWO layer indicating polycrystalline domains. Fig.~\ref{fig:dft}I--J show the corresponding FFTs from two grains identified in Fig.~\ref{fig:dft}H, showing well-defined diffraction spots consistent with a polycrystalline IWO phase.

Previous studies have reported that ALD of \ce{HfO2} on indium oxide increases the concentration of oxygen vacancy related defects inside indium oxide~\cite{si2021high}. The \ce{Hf-O} bond has a higher bond dissociation energy than the \ce{In-O} bond. Consequently, the formation of \ce{Hf-O} bonds scavenges oxygen from \ce{In-O}, generating vacancies. Interestingly, we observed that \ce{HfO2} deposition atop indium tungsten oxide transforms the initially amorphous phase into a polycrystalline phase. The observed amorphous-to-polycrystalline transition possibly arises from growth-induced stress generated during ALD of \ce{HfO2}, which promotes structural rearrangement in the underlying AOS. We performed synchrotron-based grazing-incidence X-ray diffraction (GIXRD) analysis on the channel-first sample, where a pronounced \ce{In2O3} (222) peak was observed, indicating a higher degree of crystallinity, as shown in SM fig.~S8. An increased concentration of oxygen vacancy related defects in indium oxide is also expected to lead to a slight expansion of the lattice parameter\cite{qiu2024irreversible}. A magnified STEM image for the channel-first case shows that the indium tungsten oxide film has grains with multiple orientations, as shown in SM fig.~S9. Grain boundaries associated with this polycrystalline phase can introduce defect states and, at smaller technology nodes, exacerbate device-to-device variability. Consequently, channel-first integration is ill-suited for AOS based GAA FETs. By contrast, depositing IWO after \ce{HfO2} in the channel-last case preserves a pristine, low-defect and amorphous structure. 

We further performed Density Functional Theory (DFT) calculation to understand the impact of oxygen vacancy on the electronic band structures. Fig.~\ref{fig:dft}K shows the atomistic model of the channel-first stack with oxygen vacancies inside the indium oxide. Oxygen vacancies with an areal concentration of $N_\mathrm{vac} = 1\times10^{14}\,\mathrm{cm}^{-2}$ were introduced into the indium oxide lattice. The element‑projected density of states (DOS) in Fig.~\ref{fig:dft}L exhibits finite spectral weight at the Fermi level ($E_\mathrm{F}$, set to $0~\mathrm{eV}$) dominated by In‑$5s$ states, together with three donor‑like levels at the conduction‑band (CB) edge, consistent with degenerate $n$‑type doping. The in‑plane band structure $E(k_\mathrm{x})$ (Fig.~\ref{fig:dft}M) shows that oxygen vacancies introduce donor levels that donate electrons to the CB, shifting $E_\mathrm{F}$ into the CB, indicative of degenerate n-type doping. In addition, vacancy related impurity bands appear near the CB edge and are nearly dispersionless; their weak dispersion reflects strong real‑space localization. Additional analysis with different oxygen vacancy concentrations is provided in SM fig.~S10--S16.

In contrast, in the absence of oxygen vacancies in indium oxide, the DOS places \(E_{\mathrm{F}}\) below the indium oxide CB minimum (CBM), yielding a clean gap with no states at \(E_{\mathrm{F}}\) (Fig.~\ref{fig:dft}N). A corresponding atomistic model of channel-last amorphous case is added in the SM. The corresponding \(E(k_{\mathrm{x}})\) plot (Fig.~\ref{fig:dft}O) shows only conduction sub-bands above \(E_{\mathrm{F}}\) and no flat defect bands. The curved features, correspond to band-like states, are away from the \(E_{\mathrm{F}}\), enabling the channel to be turned off. Overall, the calculations indicate that reducing the oxygen-vacancy concentration shifts \(E_{\mathrm{F}}\) away from the CB and suppresses metallic behavior~\cite{xu2015observation}. In the channel-last case, the undesired vacancy concentration remains low, leaving only delocalized, band-like states and allowing the channel to be turned off.

\section*{Discussion and Outlook}

We have presented the first demonstration of the CL GAA nanosheet FET that addresses a critical barrier in M3D integration for BEOL-compatible oxide-semiconductors. By depositing the semiconducting channel after forming the high-$\kappa$ dielectric within a released cavity, we circumvent defect formation and structural modifications that would otherwise degrade the AOS channel. This yields transistors with stable, normally-off operation, steep S.S., and high on-state current densities, all achieved without any post-fabrication treatments and in a self aligned structure. Furthermore, in situ channel deposition ensures a pristine interface, contributing to good S.S. These experimental results are corroborated by TCAD simulations, HAADF-STEM imaging, GIXRD, and DFT calculations. From a fundamental standpoint, our results answer the long standing unresolved question of why back-gated AOS devices generally exhibit good behavior, whereas top-gated or GAA AOS devices (fabricated using traditional methods) often exhibit fully-on behavior in the pristine state. Our analysis shows that a traditional channel-first process transforms the AOS channel into a polycrystalline film, thus, the channel-last approach is the most promising route to AOS GAA FETs suitable for M3D.

The channel-last paradigm offers a broadly applicable integration strategy. This approach can be generalized to any ALD-grown semiconductor channel material, from amorphous oxide semiconductors to two-dimensional materials, especially when gate dielectric deposition is chemically aggressive or detrimental. The versatility of our channel-last scheme also opens up possibilities for advanced AOS-based complementary FET architectures while maintaining a BEOL-compatible process flow without degrading the channel. This channel-last approach could be suitable for AOS‑based 3D DRAM, where high‑temperature needed to fabricate the capacitor degrades the AOS access FET. This degradation can be avoided altogether by depositing the FET channel after the fabrication of the capacitor is completed. By enabling all‑at‑once channel deposition of multilayer nanosheets for 3D memory and logic, our channel‑last approach provides a bit‑cost‑scalable fabrication pathway, similar to 3D NAND flash, which is extremely important as the fabrication energy cost of multilayer 3D chip surpass the operational energy cost. The CL GAA concept unlocks manufacturable, monolithically stacked BEOL logic and memory that break new ground in performance, density, and energy efficiency.

\begin{figure}[!h]
  \centering
  \includegraphics[width=\linewidth]{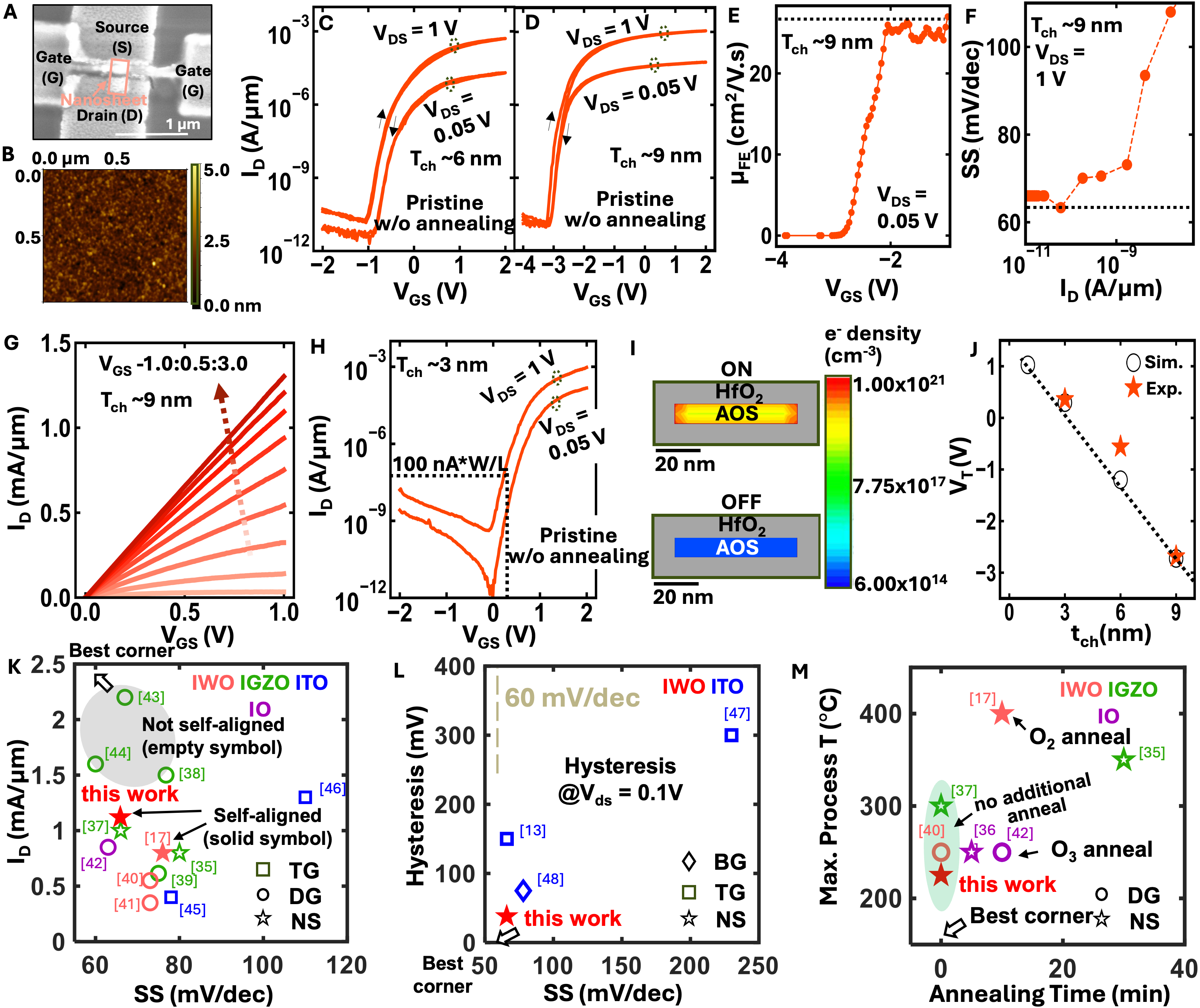}
  \caption{\textbf{Electrical performance of CL GAA FETs.} (\textbf{A}) A SEM image of a representative device structure and (\textbf{B}) an AFM image of IWO deposited on a control sample is showing uniform and smooth suface. (\textbf{C}) $I_\mathrm{D}$ versus $V_\mathrm{GS}$ characteristics for a CL GAA FET with $t_\mathrm{ch}$ = 6 nm measured at $V_\mathrm{DS} = 0.05~\text{V}$ and 1 V. (\textbf{D}) $I_\mathrm{D}$ versus $V_\mathrm{GS}$ characteristics for a CL GAA FET with $t_\mathrm{ch}$ = 9 nm, measured at $V_\mathrm{DS} = 0.05~\text{V}$ and 1 V. Dual sweep shows very low hysteresis of about 0.038 V at $V_\mathrm{DS} = 0.05~\text{V}$. (\textbf{E}) Extracted field‑effect mobility at $V_\mathrm{DS} = 0.05~\text{V}$. (\textbf{F}) Measured S.S. distribution showing a minimum value of 63.3 mV/dec, and an average over one decade is 65.9 mV/dec. (\textbf{G}) Output characteristics of $t_\mathrm{ch}$ = 9 nm device, showing a maximum current of 1 mA/µm at $V_\mathrm{DS} = 1~\text{V}$ and $V_\mathrm{GS} = 2~\text{V}$. (\textbf{H}) $I_\mathrm{D}$ versus $V_\mathrm{GS}$ for a CL GAA FET with a $t_\mathrm{ch}$ = 3 nm measured at $V_\mathrm{DS} = 0.05~\text{V}$ and 1 V.
}
  \label{fig:elec}
\end{figure}
\clearpage

\begin{figure}[!h]
    \ContinuedFloat
    \centering
    \caption{(\textbf{I}) Simulated electron density (cm\textsuperscript{-3}) distribution in the nanosheet cross-section. As expected for a GAA geometry, the electron density is high along the entire nanosheet periphery in the on state and strongly suppressed in the off state, indicating uniform depletion. The simulation was calibrated to the measured transfer characteristics of a CL GAA FET with $t_\mathrm{ch}$= 9 nm under the same bias conditions. The map shown corresponds to representative on‑state and off-state conditions. (\textbf{J}) simulated and experimental $V_\mathrm{T}$ distribution shows reasonable agreement. $ (\textbf{K}-\textbf{M})$ Benchmark analysis comparing the $I_\mathrm{on}$ (extracted at $V_\mathrm{DS} = 1~\text{V}$,$V_\mathrm{GS} = 2~\text{V}$), S.S., hysteresis and processing temperature, of AOS GAA, top‑gate, dual gate, with CL GAA FET (9 nm channel thickness). CL GAA FET achieves lower S.S., higher $I_\mathrm{on}$, , a lower hysteresis, and the lowest processing temperature. 
}

\end{figure}

\begin{figure}[!h]
  \centering
  \includegraphics[width=1\linewidth]{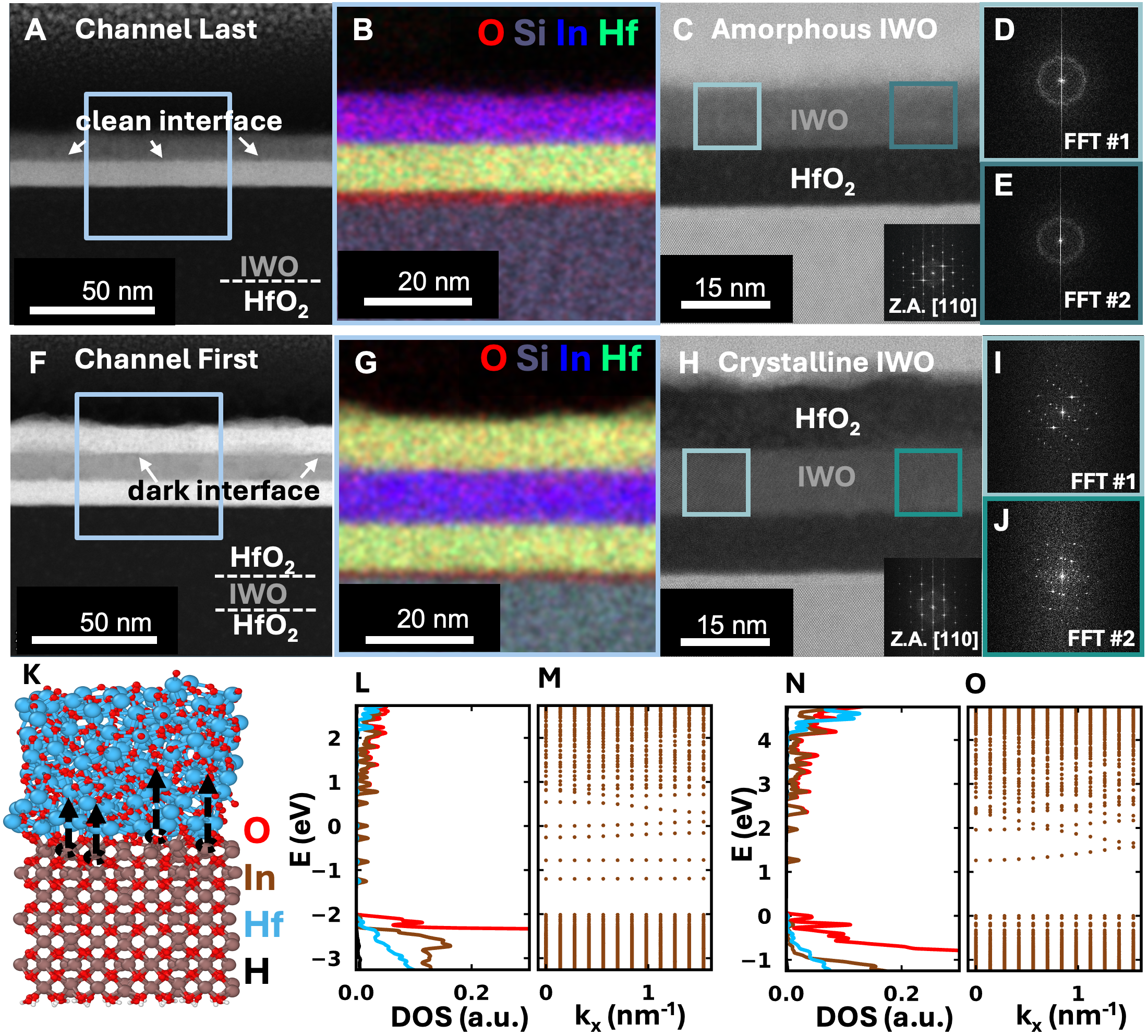}
  \caption{\textbf {Indium oxide/hafnium oxide stack analysis.} (\textbf{A}) HAADF-STEM image of IWO (low-Z-contrast)/\ce{HfO2} (high-Z-contrast) stack where \ce{HfO2} was deposited first on a blanket \ce{SiO2} substrate, followed by IWO (channel-last). The interface between IWO and \ce{HfO2} is sharp and clean (indicated by the guide arrow). (\textbf{B}) X-EDS elemental maps of O, Si, In, and Hf, confirming the expected composition and layer sequence. \textbf{(C)} High-resolution BF-STEM image along the \ce{SiO2} [110] zone axis with an inset FFT, showing the amorphous nature of the IWO layer. Selected regions of interest (ROIs) are indicated for FFT analysis. (\textbf{D}-\textbf{E}) FFTs from the ROIs in (\textbf{C}) exhibit diffuse amorphous rings without pronounced diffraction spots. The vertical streak originates from the BF-STEM scan direction. (\textbf{F}) HAADF-STEM images of a \ce{HfO2}/IWO stack where IWO is deposited first on a blanket \ce{SiO2} substrate, followed by \ce{HfO2} deposition (channel-first). Some dark spots at the interface between IWO and \ce{HfO2} are observed (indicated by the guide arrow).
}
  \label{fig:dft}
\end{figure}
\clearpage

\begin{figure}[!h]
    \ContinuedFloat
    \centering

   \caption{ (\textbf{G}) X-EDS elemental mapping confirms the chemical composition and distribution of all layers. \textbf{(H)} Lattice-resolved BF-STEM image along the \ce{SiO2} [110] zone axis (inset FFT), with ROIs in the IWO layer indicating polycrystalline domains. (\textbf{I}--\textbf{J}) Corresponding FFTs from two grains identified in (\textbf{H}), showing well-defined diffraction spots consistent with a polycrystalline IWO phase. All FFTs have a lateral dimension of 30 nm\textsuperscript{-1}. (\textbf{K}) Atomistic image of indium oxide/hafnium oxide with oxygen vacancies introduced in indium oxide ($N_\mathrm{vac}=1\times10^{14}\mathrm{cm}^{-2}$). (\textbf{L}) Element‑projected DOS versus energy plot. The $E_\mathrm{F}$ ($E_\mathrm{F}=0~\mathrm{eV}$) shifts into the conduction band (n‑type doping), and three donor‑like charged defect levels appear near the CB edge. (\textbf{M}) Energy (E) versus in‑plane wavevector $k_\mathrm{x}$ (nm$^{-1}$) dispersion plot with an oxygen vacancy. Nearly flat, vacancy‑localized donor bands appear at or just below the CB edge and cross $E_\mathrm{F}$, indicating metallic conduction. (\textbf{N}) DOS versus energy plot of indium oxide/hafnium oxide without oxygen vacancies. The $E_\mathrm{F}$ is below the indium oxide CBM, with a clean gap and no states at $E_\mathrm{F}$. (\textbf{O}) Energy versus in‑plane wavevector $k_\mathrm{x}$ (nm$^{-1}$) dispersion plot without an oxygen vacancy. The $E(k_\mathrm{x})$ plot shows only conduction subbands above $E_\mathrm{F}$ and no horizontal defect bands crossing $E_\mathrm{F}$.
}
\end{figure}

\vspace{1em}
\noindent\textbf{Materials and Methods:} Provided in the Supplementary Materials (sufficient to replicate the study). \\
\noindent\textbf{Data and Code:} All data supporting this work are included in the main text and Supplementary Materials. \\
\noindent\textbf{Acknowledgment:}
Supported in part by SRC JUMP 2.0 PRISM and CHIMES Center, Stanford Differentiated Access Memory (DAM), SystemX Alliance, Stanford NMTRI, TSMC-Stanford Joint Development Project (P.C.M.). Part of this work was performed at Nano at Stanford (RRID SCR 026695). Authors acknowledge help and support for the AFM measurements from Christina Newcomb. Stanford authors thank MSS USA Corp. (Sunnyvale, CA, USA) for high-quality TEM sample preparation and examination and Dr. David Fried of Lam Research for providing access to  Coventor SEMulator3D for process simulation. Use of the Stanford Synchrotron Radiation Lightsource at SLAC National Accelerator Laboratory is supported by the U.S. Department of Energy, Office of Science, Office of Basic Energy Sciences under Contract No. DE-AC02-76SF00515. Y.C. acknowledge the support from the Japan Society for the Promotion of Science (JSPS) overseas research fellowship. J.A.D., A.M.G., and P.M. acknowledge the financial support from the U.S. Department of Energy Office of Science National Quantum Information Science Research Centers as part of the Q-NEXT center. Y.C. and F.F.A. would like to thank the support from the Stanford Energy Postdoctoral Fellowship, and Precourt Institute for Energy.\\

\clearpage

\clearpage
\bibliographystyle{science}
\bibliography{scibib}

@misc{IRDS2022,
  title        = {International Roadmap for Devices and Systems (IRDS): Executive Summary},
  author       = {{IEEE IRDS}},
  year         = {2022},
  howpublished = {\url{https://irds.ieee.org/roadmap}}
}

@article{kim2005off,
  title={Off-state current and performance analysis for double-gate CMOS with non-self-aligned back gate},
  author={Kim, Keunwoo and Hanafi, Hussein I and Cai, Jin and Chuang, Ching-Te},
  journal={IEEE transactions on electron devices},
  volume={52},
  number={9},
  pages={2104--2107},
  year={2005},
  publisher={IEEE}
}

@article{qiu2024irreversible,
  title={Irreversible Lattice Expansion Effects in Nanoscale Indium Oxide for CO2 Hydrogenation Catalysis},
  author={Qiu, Chenyue and Sun, Junchuan and Li, Mengsha and Mao, Chengliang and Song, Rui and Zhang, Zeshu and Perovic, Doug D and Howe, Jane Y and Wang, Lu and Ozin, Geoffrey A},
  journal={Journal of the American Chemical Society},
  volume={146},
  number={49},
  pages={33997--34007},
  year={2024},
  publisher={ACS Publications}
}

@article{si2021high,
  title={High-performance atomic-layer-deposited indium oxide 3-D transistors and integrated circuits for monolithic 3-D integration},
  author={Si, Mengwei and Lin, Zehao and Chen, Zhizhong and Ye, Peide D},
  journal={IEEE Transactions on Electron Devices},
  volume={68},
  number={12},
  pages={6605--6609},
  year={2021},
  publisher={IEEE}
}

@article{jonsson2018self,
  title={A self-aligned gate-last process applied to all-III--V CMOS on Si},
  author={J{\"o}nsson, Adam and Svensson, Johannes and Wernersson, Lars-Erik},
  journal={IEEE Electron Device Letters},
  volume={39},
  number={7},
  pages={935--938},
  year={2018},
  publisher={IEEE}
}

@article{li2021first,
  title={First demonstration of novel vertical gate-all-around field-effect-transistors featured by self-aligned and replaced high-$\kappa$ metal gates},
  author={Li, Chen and Zhu, Huilong and Zhang, Yongkui and Wang, Qi and Yin, Xiaogen and Li, Junjie and Wang, Guilei and Kong, Zhenzhen and Ai, Xuezheng and Xie, Lu and others},
  journal={Nano letters},
  volume={21},
  number={11},
  pages={4730--4737},
  year={2021},
  publisher={ACS Publications}
}

@inproceedings{kim2005extremely,
  title={Extremely-scaled double-gate CMOS with non-self-aligned back gate},
  author={Kim, Keunwoo and Hanafi, Hussein I and Cai, Jin and Chuang, Ching-Te},
  booktitle={IEEE VLSI-TSA International Symposium on VLSI Technology, 2005.(VLSI-TSA-Tech).},
  pages={110--111},
  year={2005},
  organization={IEEE}
}

@inproceedings{athena2024first,
  title={First demonstration of an np oxide semiconductor complementary gain cell memory},
  author={Athena, FF and Ambrosi, E and Jana, K and Wu, CH and Hartanto, J and Lee, YM and Kuo, CC and Liu, S and Saini, B and Wang, CC and others},
  booktitle={2024 IEEE International Electron Devices Meeting (IEDM)},
  pages={1--4},
  year={2024},
  organization={IEEE}
}

@misc{IRDS2023MM,
  title        = {IRDS 2023: More Moore and System Technology Co-Optimization},
  author       = {{IEEE IRDS}},
  year         = {2023},
  howpublished = {\url{https://irds.ieee.org/editions/2023}}
}

@article{woods2024death,
  title={The Death of Moore’s Law: What it means and what might fill the gap going forward},
  author={Woods, Audrey},
  journal={MIT CSAIL},
  year={2024}
}

@article{franklin2022carbon,
  title={Carbon nanotube transistors: Making electronics from molecules},
  author={Franklin, Aaron D and Hersam, Mark C and Wong, H-S Philip},
  journal={Science},
  volume={378},
  number={6621},
  pages={726--732},
  year={2022},
  publisher={American Association for the Advancement of Science}
}

@article{xu2015observation,
  title={Observation of Fermi arc surface states in a topological metal},
  author={Xu, Su-Yang and Liu, Chang and Kushwaha, Satya K and Sankar, Raman and Krizan, Jason W and Belopolski, Ilya and Neupane, Madhab and Bian, Guang and Alidoust, Nasser and Chang, Tay-Rong and others},
  journal={Science},
  volume={347},
  number={6219},
  pages={294--298},
  year={2015},
  publisher={American Association for the Advancement of Science}
}

@article{lyu1993determination,
  title={Determination of the interface trap density in metal oxide semiconductor field-effect transistor through subthreshold slope measurement},
  author={Lyu, Jong-Son and Lee, Kee-Soo Nam},
  journal={Japanese journal of applied physics},
  volume={32},
  number={10R},
  pages={4393},
  year={1993},
  publisher={IOP Publishing}
}

@article{jiang2019printed,
  title={Printed subthreshold organic transistors operating at high gain and ultralow power},
  author={Jiang, Chen and Choi, Hyung Woo and Cheng, Xiang and Ma, Hanbin and Hasko, David and Nathan, Arokia},
  journal={Science},
  volume={363},
  number={6428},
  pages={719--723},
  year={2019},
  publisher={American Association for the Advancement of Science}
}

@article{lee2016subthreshold,
  title={Subthreshold Schottky-barrier thin-film transistors with ultralow power and high intrinsic gain},
  author={Lee, Sungsik and Nathan, Arokia},
  journal={Science},
  volume={354},
  number={6310},
  pages={302--304},
  year={2016},
  publisher={American Association for the Advancement of Science}
}

@article{elers2006film,
  title={Film uniformity in atomic layer deposition},
  author={Elers, K-E and Blomberg, Tom and Peussa, Marko and Aitchison, Brad and Haukka, Suvi and Marcus, Steven},
  journal={Chemical Vapor Deposition},
  volume={12},
  number={1},
  pages={13--24},
  year={2006},
  publisher={Wiley Online Library}
}

@inproceedings{lin2024enhancement,
  title={Enhancement-Mode Atomic Layer Deposited W-Doped In 2 O 3 Transistor at 55 nm Channel Length by Oxide Capping Layer with Improved Stability},
  author={Lin, Qing and Safron, Nathaniel and Zhong, Donglai and Arutchelvan, Goutham and Gilardi, Carlo and Yoo, Chanyoung and Hartanto, Jonathan and Saini, Balreen and Lai, Sheng-Chih and Pitner, Gregory and others},
  booktitle={2024 IEEE International Electron Devices Meeting (IEDM)},
  pages={1--4},
  year={2024},
  organization={IEEE}
}

@article{yoo2024atomic,
  title={Atomic layer deposition of WO3-doped In2O3 for reliable and scalable BEOL-compatible transistors},
  author={Yoo, Chanyoung and Hartanto, Jonathan and Saini, Balreen and Tsai, Wilman and Thampy, Vivek and Niavol, Somayeh Saadat and Meng, Andrew C and McIntyre, Paul C},
  journal={Nano Letters},
  volume={24},
  number={19},
  pages={5737--5745},
  year={2024},
  publisher={ACS Publications}
}

@article{lukens2018self,
  title={Self-aligned process for selectively etched p-GaN-gated AlGaN/GaN-on-Si HFETs},
  author={L{\"u}kens, Gerrit and Hahn, Herwig and Kalisch, Holger and Vescan, Andrei},
  journal={IEEE Transactions on Electron Devices},
  volume={65},
  number={9},
  pages={3732--3738},
  year={2018},
  publisher={IEEE}
}

@article{liang1994self,
  title={Self-aligned dry-etching process for waveguide diode ring lasers},
  author={Liang, James J and Ballantyne, Joseph M},
  journal={Journal of Vacuum Science \& Technology B: Microelectronics and Nanometer Structures Processing, Measurement, and Phenomena},
  volume={12},
  number={5},
  pages={2929--2932},
  year={1994},
  publisher={American Vacuum Society}
}

@article{lee1997dry,
  title={Dry release for surface micromachining with HF vapor-phase etching},
  author={Lee, Y-I and Park, K-H and Lee, Jonghyun and Lee, C-S and Yoo, Hyung Joun and Kim, C-J and Yoon, Y-S},
  journal={Journal of microelectromechanical systems},
  volume={6},
  number={3},
  pages={226--233},
  year={1997},
  publisher={IEEE}
}

@inproceedings{witvrouw2000comparison,
  title={Comparison between wet HF etching and vapor HF etching for sacrificial oxide removal},
  author={Witvrouw, Ann and Du Bois, Bert and De Moor, Piet and Verbist, Agnes and Van Hoof, Chris A and Bender, Hugo and Baert, Christiaan},
  booktitle={Micromachining and Microfabrication Process Technology VI},
  volume={4174},
  pages={130--141},
  year={2000},
  organization={SPIE}
}

@article{zywotko2017thermal,
  title={Thermal atomic layer etching of ZnO by a “Conversion-Etch” mechanism using sequential exposures of hydrogen fluoride and trimethylaluminum},
  author={Zywotko, David R and George, Steven M},
  journal={Chemistry of Materials},
  volume={29},
  number={3},
  pages={1183--1191},
  year={2017},
  publisher={ACS Publications}
}

@inproceedings{jana2025key,
  title={Key to Low Supply Voltage: Transition Region of Oxide Semiconductor Transistors},
  author={Jana, K and Kang, J and Liu, S and Athena, FF and Huang, C-H and Tang, Y and Chen, HJ-Y and Saini, B and Hartanto, J and Bennett, RKA and others},
  booktitle={2025 Symposium on VLSI Technology and Circuits (VLSI Technology and Circuits)},
  pages={1--3},
  year={2025},
  organization={IEEE}
}

@incollection{tsuya1994oxygen,
  title={Oxygen effect on electronic device performance},
  author={Tsuya, H},
  booktitle={Semiconductors and Semimetals},
  volume={42},
  pages={619--667},
  year={1994},
  publisher={Elsevier}
}

@article{kelly2018annealing,
  title={Annealing and impurity effects in Co thin films for MOL contact and BEOL metallization},
  author={Kelly, James and Kamineni, Vimal and Lin, X and Pacquette, A and Hopstaken, Marinus and Liang, Yong and Amanapu, Hari and Peethala, B and Jiang, Liying and Demarest, J and others},
  journal={Journal of the Electrochemical Society},
  volume={166},
  number={1},
  pages={D3100},
  year={2018},
  publisher={IOP Publishing}
}

@inproceedings{sarkar2024first,
  title={First Demonstration of W-Doped In 2 O 3 Gate-All-Around (GAA) Nanosheet FET with Improved Performance and Record Threshold Voltage Stability},
  author={Sarkar, Eknath and Zhang, Chengyang and Chakraborty, Dyutimoy and Waqar, Faaiq G and Kirtania, Sharadindugopal and Aabrar, Khandker Akif and Park, Hyeonwoo and Shin, Jaewon and Tian, Mengkun and Khan, Asif I and others},
  booktitle={2024 IEEE International Electron Devices Meeting (IEDM)},
  pages={1--4},
  year={2024},
  organization={IEEE}
}

@article{fortunato2012oxide,
  title={Oxide semiconductor thin-film transistors: a review of recent advances},
  author={Fortunato, Elvira and Barquinha, Pedro and Martins, Rodrigo},
  journal={Advanced materials},
  volume={24},
  number={22},
  pages={2945--2986},
  year={2012},
  publisher={Wiley Online Library}
}

@article{yan2025clean,
  title={A clean van der Waals interface between the high-k dielectric zirconium oxide and two-dimensional molybdenum disulfide},
  author={Yan, Han and Wang, Yan and Li, Yang and Phuyal, Dibya and Liu, Lixin and Guo, Hailing and Guo, Yuzheng and Lee, Tien-Lin and Kim, Minhyuk and Jeong, Hu Young and others},
  journal={Nature Electronics},
  pages={1--7},
  year={2025},
  publisher={Nature Publishing Group UK London}
}

@inproceedings{athena2025orthogonal,
  title={Orthogonal V\textsubscript{T} Tuning for Oxide Semiconductor 2T Gain Cell Enabled by Interface Dipole Engineering},
  author={Athena, FF and Kang, J and Passlack, M and Safron, N and Dede, D and Jana, K and Saini, B and Wang, X and Liu, S and Hartanto, J and others},
  booktitle={2025 Symposium on VLSI Technology and Circuits (VLSI Technology and Circuits)},
  pages={1--3},
  year={2025},
  organization={IEEE}
}

@inproceedings{Mallik2017IEDM,
  author    = {Mallik, Arindam and Vandooren, Anne and Witters, Liesbeth and Walke, Alexander and Franco, Jan and Sherazi, Yasir and Weckx, Philippe and Yakimets, Dmytro and Bardon, Muriel and Parvais, Benoit and Debacker, Pieter and Ku, Bo Wu and Lim, Seung K. and Mocuta, Andrei and Mocuta, Dan and Ryckaert, Jan and Collaert, Nicolas and Raghavan, Praveen},
  title     = {The Impact of Sequential-3D Integration on Semiconductor Scaling Roadmap},
  booktitle = {2017 IEEE International Electron Devices Meeting (IEDM)},
  year      = {2017},
  pages     = {32.1.1--32.1.4},
  doi       = {10.1109/IEDM.2017.8268483}
}

@article{Si2021TED_Emode,
  author  = {Si, Mengwei and Charnas, Adam and Lin, Zehao and Ye, Peide D.},
  title   = {Enhancement-Mode ALD In$_2$O$_3$ Transistors with Maximum Drain Current of 2.2 A/mm at $V_{DS}=0.7$ V},
  journal = {IEEE Transactions on Electron Devices},
  year    = {2021},
  volume  = {68},
  number  = {3},
  pages   = {1075--1080},
  doi     = {10.1109/TED.2021.3053229}
}

@article{Wahid2023EDL,
  author  = {Wahid, Sumaiya and Kwon, Jimin and Qin, Shengjun and Ko, Jung-Soo and Wong, H.-S. Philip and Pop, Eric},
  title   = {Effect of Top-Gate Dielectric Deposition on the Performance of Indium Tin Oxide Transistors},
  journal = {IEEE Electron Device Letters},
  year    = {2023},
  volume  = {44},
  number  = {6},
  pages   = {951--954},
  doi     = {10.1109/LED.2023.3265316}
}

@article{sarkar2025first,
  title={First Demonstration of High-Performance and Extremely Stable W-Doped In 2 O 3 Gate-All-Around (GAA) Nanosheet FET},
  author={Sarkar, Eknath and Zhang, Chengyang and Chakraborty, Dyutimoy and Kirtania, Sharadindu Gopal and Aabrar, Khandker Akif and Park, Hyeonwoo and Shin, Jaewon and Lee, Hyun Jae and Tian, Mengkun and Khan, Asif I and others},
  journal={IEEE Transactions on Electron Devices},
  year={2025},
  publisher={IEEE}
}

@article{Charnas2024ADMA,
  author  = {Charnas, Adam and Zhang, Jingxuan and Si, Mengwei and Ye, Peide D.},
  title   = {Review—Extremely Thin Amorphous Indium Oxide Transistors},
  journal = {Advanced Materials},
  year    = {2024},
  volume  = {36},
  number  = {9},
  pages   = {2304044},
  doi     = {10.1002/adma.202304044}
}

@inproceedings{franke2017epe,
  title={EPE analysis of sub-N10 BEOL flow with and without fully self-aligned via using Coventor SEMulator3D},
  author={Franke, Joern-Holger and Gallagher, Matt and Murdoch, Gayle and Halder, Sandip and Juncker, Aurelie and Clark, William},
  booktitle={Metrology, Inspection, and Process Control for Microlithography XXXI},
  volume={10145},
  pages={670--679},
  year={2017},
  organization={SPIE}
}

@INPROCEEDINGS{han2021first,
  author={Han, Kaizhen and Kong, Qiwen and Kang, Yuye and Sun, Chen and Wang, Chengkuan and Zhang, Jishen and Xu, Haiwen and Samanta, Subhranu and Zhou, Jiuren and Wang, Haibo and Thean, Aaron Voon-Yew and Gong, Xiao},
  booktitle={2021 Symposium on VLSI Technology}, 
  title={First Demonstration of Oxide Semiconductor Nanowire Transistors: a Novel Digital Etch Technique, IGZO Channel, Nanowire Width Down to ~20 nm, and Ion Exceeding 1300 μA/μm}, 
  year={2021},
  volume={},
  number={},
  pages={1-2},
  doi={}}

@ARTICLE{zhang2022a,
  author={Zhang, Zhuocheng and Lin, Zehao and Liao, Pai-Ying and Askarpour, Vahid and Dou, Hongyi and Shang, Zhongxia and Charnas, Adam and Si, Mengwei and Alajlouni, Sami and Shakouri, Ali and Wang, Haiyan and Lundstrom, Mark and Maassen, Jesse and Ye, Peide D.},
  journal={IEEE Electron Device Letters}, 
  title={A Gate-All-Around inO Nanoribbon FET With Near 20 mA/m Drain Current}, 
  year={2022},
  volume={43},
  number={11},
  pages={1905-1908},
  doi={10.1109/LED.2022.3210005}}

@INPROCEEDINGS{li2023first,
  author={Li, Qijun and Zhao, Wenjie and Hu, Qianlan and Gu, Chengru and Zhu, Shenwu and Liu, Honggang and Huang, Ru and Wu, Yanqing},
  booktitle={2023 International Electron Devices Meeting (IEDM)}, 
  title={First Demonstration of Sequential Integration for Stacked Gate-All-Around a-IGZO Nanosheet Transistors with Record Id = 2.05 mA/µm, gm = 1.13 mS/µm and Ultralow SS = 66 mV/dec}, 
  year={2023},
  volume={},
  number={},
  pages={1-4},
  doi={10.1109/IEDM45741.2023.10413891}}

@INPROCEEDINGS{lu2022performance,
  author={Lu, Wendong and Zhu, Zhengyong and Chen, Kaifei and Liu, Menggan and Kang, Bok-Moon and Duan, Xinlv and Niu, Jiebin and Liao, Fuxi and Dan, Wang and Wu, Xie-Shuai and Son, Joohwan and Xiao, De-Yuan and Wang, Gui-Lei and Yoo, Abraham and Cao, Kan-Yu and Geng, Di and Lu, Nianduan and Yang, Guanhua and Zhao, Chao and Li, Ling and Liu, Ming},
  booktitle={2022 International Electron Devices Meeting (IEDM)}, 
  title={First Demonstration of Dual-Gate IGZO 2T0C DRAM with Novel Read Operation, One Bit Line in Single Cell, ION=1500 μA/μm@VDS=1V and Retention Time>300s}, 
  year={2022},
  volume={},
  number={},
  pages={26.4.1-26.4.4},
  doi={10.1109/IEDM45625.2022.10019488}}

@INPROCEEDINGS{chen2022scaling,
  author={Chen, Kaifei and Niu, Jiebin and Yang, Guanhua and Liu, Menggan and Lu, Wendong and Liao, Fuxi and Huang, Kailiang and Duan, XinLv and Lu, Congyan and Wang, Jiawei and Wang, Lingfei and Li, Mengmeng and Geng, Di and Zhao, Chao and Wang, Guilei and Lu, Nianduan and Li, Ling and Liu, Ming},
  booktitle={2022 IEEE Symposium on VLSI Technology and Circuits (VLSI Technology and Circuits)}, 
  title={Scaling Dual-Gate Ultra-thin a-IGZO FET to 30 nm Channel Length with Record-high Gm,max of 559 µS/µm at VDS=1 V, Record-low DIBL of 10 mV/V and Nearly Ideal SS of 63 mV/dec}, 
  year={2022},
  volume={},
  number={},
  pages={298-299},
  doi={10.1109/VLSITechnologyandCir46769.2022.9830389}}

@INPROCEEDINGS{ye2020double,
  author={Ye, H. and Gomez, J. and Chakraborty, W. and Spetalnick, S. and Dutta, S. and Ni, K. and Raychowdhury, A. and Datta, S.},
  booktitle={2020 IEEE International Electron Devices Meeting (IEDM)}, 
  title={Double-Gate W-Doped Scaling Dual-Gate Ultra-thin a-IGZO FET to 30 nm Channelfor Monolithic 3D Capacitorless Gain Cell eDRAM}, 
  year={2020},
  volume={},
  number={},
  pages={28.3.1-28.3.4},
  doi={10.1109/IEDM13553.2020.9371981}}

@INPROCEEDINGS{chakraborty2020beol,
  author={Chakraborty, Wriddhi and Grisafe, Benjamin and Ye, Huacheng and Lightcap, Ian and Ni, Kai and Datta, Suman},
  booktitle={2020 IEEE Symposium on VLSI Technology}, 
  title={BEOL Compatible Dual-Gate Ultra Thin-Body W-Doped Indium-Oxide Transistor with Ion = 370μA/μm, SS = 73mV/dec and Ion /Ioff Ratio > 4×109}, 
  year={2020},
  volume={},
  number={},
  pages={1-2},
  doi={10.1109/VLSITechnology18217.2020.9265064}}

@article{ye2025high,
  author = {Ye, Chenchen and Li, Jiakang and Hong, Peiyan and Zhao, Jiaming and Miao, Xiangshui and Li, Xuefei},
  title = {High-Performance Atomic-Layer-Deposited Dual-Gate InGaO Thin-Film Transistors},
  journal = {Nano Letters},
  volume = {25},
  number = {21},
  pages = {8541-8546},
  year = {2025},
  doi = {10.1021/acs.nanolett.5c01059},
  note ={PMID: 40371998}}

@INPROCEEDINGS{tong2025boosted,
  author={Tong, Anyu and Hu, Qianlan and Zeng, Min and Zhu, Yuzhe and Zhao, Wenjie and Wang, Zhiyu and Wu, Yanqing},
  booktitle={2025 9th IEEE Electron Devices Technology \& Manufacturing Conference (EDTM)}, 
  title={Boosted Performance of Atomic-Layer-Deposited Dual-Gate Indium-Gallium-Zinc-Oxide Transistors}, 
  year={2025},
  volume={},
  number={},
  pages={1-3},
  doi={10.1109/EDTM61175.2025.11040472}}

@INPROCEEDINGS{zhao2024first,
  author={Zhao, Wenjie and Zhu, Shenwu and Li, Qijun and Hu, Qianlan and Liu, Honggang and Tong, Anyu and Zeng, Min and Huang, Ru and Wu, Yanqing},
  booktitle={2024 IEEE International Electron Devices Meeting (IEDM)}, 
  title={First Demonstration of Double-Gate IGZO Transistors with Ideal Subthreshold Swing of 60 mV/dec at Room Temperature and 76 mV/dec at 380 K Over 5 Decades and gm Exceeding 1 mS/µm with Contact Length Scaling}, 
  year={2024},
  volume={},
  number={},
  pages={1-4},
  doi={10.1109/IEDM50854.2024.10873547}}

@ARTICLE{yang2025performance,
  author={Yang, Linlong and Chen, Xi and Ma, Lan and Luo, Binbin and Yang, Ming and Meng, Wei and Teng, Jiahui and Zhu, Bao and Ding, Shi-Jin and Wu, Xiaohan},
  journal={IEEE Electron Device Letters}, 
  title={Performance Optimization of Short-Channel Top-Gate Ultrathin InSnO Thin-Film Transistor With Gate Dielectric Engineering}, 
  year={2025},
  volume={46},
  number={11},
  pages={2054-2057},
  doi={10.1109/LED.2025.3606654}}

@ARTICLE{gu2023high,
  author={Gu, Chengru and Hu, Qianlan and Zhu, Shenwu and Li, Qijun and Zeng, Min and Liu, Honggang and Kang, Jiyang and Liu, Shiyuan and Wu, Yanqing},
  journal={IEEE Electron Device Letters}, 
  title={High-Performance Short-Channel Top-Gate Indium-Tin-Oxide Transistors by Optimized Gate Dielectric}, 
  year={2023},
  volume={44},
  number={5},
  pages={837-840},
  doi={10.1109/LED.2023.3262684}}

@ARTICLE{luo2025the,
  author={Luo, Jie and Bao, Yunjiao and Yang, Yanyu and Lu, Yupeng and Wang, Guilei and Xu, Gaobo and Yin, Huaxiang and Zhao, Chao and Luo, Jun},
  journal={IEEE Transactions on Electron Devices}, 
  title={The Investigation of the Hysteresis and Reliability Mechanism of Amorphous Oxide Semiconductor Thin-Film Transistors Applied in Dynamic Random Access Memory}, 
  year={2025},
  volume={72},
  number={4},
  pages={1763-1768},
  doi={10.1109/TED.2025.3545007}}
\end{document}